\documentstyle{article}
\makeatletter
\def\section{\@startsection {section}{1}{\z@}{-3.5ex plus -1ex minus
 -.2ex}{2.3ex plus .2ex}{\large\bf}}
\def\subsection{\@startsection{subsection}{2}{\z@}{-3.25ex plus -1ex minus
 -.2ex}{1.5ex plus .2ex}{\bf}}
\long\def\@caption#1[#2]#3{\par\addcontentsline{\csname
  ext@#1\endcsname}{#1}{\protect\numberline{\csname
  the#1\endcsname}{\ignorespaces #2}}\begingroup
    \@parboxrestore
    \small\bf
    \@makecaption{\csname fnum@#1\endcsname}{\ignorespaces #3}\par
  \endgroup}
\def\fnum@figure{Fig.\ \thefigure}
\makeatother

\renewcommand{\title}[1]{\mbox{\ }\vspace*{5pt}
			\setcounter{footnote}{0}
			\renewcommand{\thefootnote}{\fnsymbol{footnote}}
			\begin{center}
			{\Large \bf #1\\ \ \\}
			R.J. van Glabbeek\makebox[0pt][l]{\footnotemark}\\
			\footnotesize
			Computer Science Department, Stanford University\\
			Stanford, CA 94305-9045, USA.\\
			{\tt rvg@cs.stanford.edu}
			\end{center}
			\footnotetext{This work was supported by ONR under
			grant number N00014-92-J-1974.}
			\renewcommand{\thefootnote}{\arabic{footnote}}
			\setcounter{footnote}{0}
			\vspace{5pt}			}
\renewenvironment{abstract}{\begin{list}{}
			{\leftmargin 8mm
			\rightmargin\leftmargin
			\listparindent 1.5em
			\parsep 0pt plus 1pt}
			\small\item}{\end{list}}

\newcommand{\eg}{{\rm e.g.}}

\newcommand{\mathify}[1]{\ifmmode{#1}\else\mbox{$#1$}\fi}
\newcommand{\proves}{\vdash}

\let\phi=\varphi

\newtheorem{theorem}{Theorem}[section]
\newtheorem{lemma}[theorem]{Lemma}
\newtheorem{proposition}[theorem]{Proposition}
\newtheorem{definition}[theorem]{Definition}
\newtheorem{corollary}[theorem]{Corollary}

\newenvironment{proof}{{\bf Proof:}\small}{\mbox{}\unskip~~\hfill$\Box$\medskip}

\newcommand{\bigger}[3]{\setbox0=\hbox{$#3$}\ht0=1.05\ht0\mathify{#1\box0#2}}

\def\set#1{\bigger{\left\{}{\right\}}{#1}}
\def\parens#1{\bigger{\left(}{\right)}{#1}}

\newcommand{\Bigger}[3]{\setbox0=\hbox{$#3$}\ht0=1.10\ht0\mathify{#1\box0#2}}

\def\Parens#1{\Bigger{\left(}{\right)}{#1}}

\newcommand{\mv}[1]{\mathrel{\stackrel{#1}{\rightarrow}}}
\newcommand{\mvt}[1]{\mathrel{\stackrel{#1}{\Rightarrow}}}
\newcommand{\nv}[1]{\mathrel{\stackrel{#1}{\not\rightarrow}}}

\newcommand{\TwoRules}[4]{
\begin{displaymath}
\begin{array}{c}
#1 \\\hline 
#2
\end{array}
\qquad
\begin{array}{c}
#3 \\\hline 
#4
\end{array}
\end{displaymath}
}

\newcommand{\ThreeRules}[6]{
\begin{displaymath}
\begin{array}{c}
#1 \\\hline 
#2
\end{array}
\qquad
\begin{array}{c}
#3 \\\hline 
#4
\end{array}
\qquad
\begin{array}{c}
#5 \\\hline 
#6
\end{array}
\end{displaymath}
}

\newcommand{\FourRules}[8]{
\begin{displaymath}
\begin{array}{c}
#1 \\\hline
#2
\end{array}
\qquad
\begin{array}{c}
#3 \\\hline
#4
\end{array}
\qquad
\begin{array}{c}
#5 \\\hline
#6
\end{array}
\qquad
\begin{array}{c}
#7 \\\hline
#8
\end{array}
\end{displaymath}
}

\newcommand{\Rel}[1]{\mathrel{\cal #1}} 
\newcommand{\Var}{\mbox{\sf Var}}
\newcommand{\Vardef}{\mbox{\em $\Var$}}
\newcommand{\deq}{\mathrel{\stackrel{\mbox{\tiny def}}{=}}} 
\newcommand{\der}[1]{\mbox{\sf der}\parens{#1}}
\newcommand{\pder}[1]{\mbox{\sf pder}\parens{#1}}
\newcommand{\LMPA}{\mbox{\sc BCCS}^{\scriptstyle f*}}
\newcommand{\OpenT}{
                   \setlength{\unitlength}{1ex}
                   \begin{picture}(1.25,1.45)
                   \put(0.4,0){\line(0,1){1.45}}
                   \put(0.85,0){\line(0,1){1.45}}
                   \put(-0.1,1.5){\line(1,0){1.45}}
                   \put(0,1.45){\line(1,0){1.25}}
                   \put(0.4,0){\line(1,0){0.45}}
                   \end{picture}}
\newcommand{\OMPA}{\OpenT}
\newcommand{\OMPAdef}{\OMPA}
\newcommand{\LMPAdef}{\LMPA}
\newcommand{\SF}{\mbox{\sc SF}}
\newcommand{\bis}[2]{ \;                                
        \raisebox{.3ex}{$\underline{\makebox[.7em]{$\leftrightarrow$}}$}
                  \,_{#1}^{#2}\,}
\newcommand{\nobis}[2]{\mbox{$\,\not\hspace{-2.5pt}     
        \raisebox{.3ex}{$\underline{\makebox[.7em]{$\leftrightarrow$}}$}
                  \,_{#1}^{#2}\,$}}
\newcommand{\obseq}{\bis{w}{}} 
\newcommand{\obscong}{\bis{w}{c}} 
\newcommand{\braeq}{\bis{b}{}} 
\newcommand{\bracong}{\bis{b}{c}} 
\newcommand{\deleq}{\bis{d}{}} 
\newcommand{\delcong}{\bis{d}{c}} 
\newcommand{\etaeq}{\bis{\eta}{}}
\newcommand{\etacong}{\bis{\eta}{c}} 
\newcommand{\ACeq}{=}
\newcommand{\bran}[1]{\underline{#1}}
\newcommand{\wild}{\aleph}

\begin{document}

\title{Axiomatizing Flat Iteration}

\begin{abstract}
Flat iteration is a variation on the original binary version of the
Kleene star operation $P^*Q$, obtained by restricting the first
argument to be a sum of atomic actions. It generalizes prefix
iteration, in which the first argument is a single action. Complete
finite equational axiomatizations are given for five notions of bisimulation
congruence over basic CCS with flat iteration, viz.~strong
congruence, branching congruence, $\eta$-congruence, delay congruence
and weak congruence. Such axiomatizations were already known for
prefix iteration and are known not to exist for general iteration.
The use of flat iteration has two main advantages over prefix iteration:
\begin{enumerate}
\item The current axiomatizations generalize to full CCS,
whereas the prefix iteration approach does not allow an elimination
theorem for an asynchronous parallel composition operator.
\item The greater expressiveness of flat iteration allows for much
shorter completeness proofs.
\end{enumerate}
In the setting of prefix iteration, the most convenient way to obtain
the completeness theorems for $\eta$-, delay, and weak congruence was
by reduction to the completeness theorem for branching congruence. In
the case of weak congruence this turned out to be much simpler than
the only direct proof found. In the setting of flat iteration on the
other hand, the completeness theorems for delay and weak (but not
$\eta$-) congruence can equally well be obtained by reduction to
the one for strong congruence, without using branching congruence as
an intermediate step.  Moreover, the completeness results for prefix
iteration can be retrieved from those for flat iteration, thus
obtaining a second indirect approach for proving completeness for
delay and weak congruence in the setting of prefix iteration.
\end{abstract}

\section{Introduction}\label{Sect:intro}

The research literature on process theory has recently witnessed a
resurgence of interest in Kleene star-like operations
\cite{BBP94,FZ94,Fok94,Sew94,CDL95,AI95,Fok95,AG95,AFGI95}.  In
\cite{CDL95} tree-based models for theories involving Kleene's star
operation $^*$ \cite{Kleene56} are studied.  \cite{BBP94} investigates the
expressive power of variations on standard process description
languages in which infinite behaviours are defined by means of $^*$
rather than by means of systems of recursion equations. The papers
\cite{FZ94,Sew94,Fok94,AI95,Fok95,AG95,AFGI95} study the possibility
of giving finite equational axiomatizations of bisimulation-like
equivalences over fragments of such languages. This study is continued
here.

In \cite{FZ94} a complete finite equational axiomatization of strong
bisimulation equivalence was given for a process algebra featuring
choice, sequential composition, and the original binary version of the
Kleene star operation $P^*Q$ \cite{Kleene56}.  \cite{Sew94} shows that
such an axiomatization does not exists in the presence of the process
0 denoting inaction, or a process denoting successful termination. The
same proof strategy can be adopted to conclude that there is no finite
equational axiomatization for weak or branching bisimulation over an
enrichment of this basic process algebra with an internal action.

For this reason restrictions of the Kleene star have been
investigated. \cite{Fok94} presents a finite, complete equational
axiomatization of strong bisimulation equivalence for Basic CCS
(the fragment of Milner's CCS \cite{Mi89} containing the
operations needed to express finite synchronization trees) with {\em
prefix iteration}.  Prefix iteration is a variation on the binary
Kleene star operation $P^*Q$,
obtained by restricting the first argument to be an atomic action.
The same is done in \cite{AG95} for {\em string iteration}.  The work
of \cite{Fok94} has been extended in \cite{AFGI95} and its
predecessors \cite{AI95,Fok95,vG95b} to cope with {\em weak}, {\em
delay}, {\em eta-} and {\em branching bisimulation congruence} in a
setting with the unobservable action $\tau$. Motivation and background
material on these behavioural congruences can be found, e.g., in
\cite{Mi89} and \cite{GW96}.  The strategy adopted in \cite{AFGI95} in
establishing the completeness results is based upon the use of
branching equivalence in the analysis of weak, delay and
$\eta$-equivalence, advocated in \cite{vG95b}. Following
\cite{vG95b}, complete axiomatizations for weak, delay and
$\eta$-congruence were obtained from one for branching congruence
by:
\begin{enumerate}
\item identifying a collection of process terms on which branching congruence
  coincides with the congruence one aims at axiomatizing, and 
\item finding an axiom system that allows for the reduction of every process 
  term to one of the required form.
\end{enumerate} 
Perhaps surprisingly, the proof for weak congruence so obtained is
simpler than the one given in \cite{AI95} which only uses properties
of weak congruence.  The direct proof method employed in \cite{AI95}
yields a long proof with many case distinctions, while the indirect
proof via branching congruence in \cite{AFGI95} is considerably
shorter, and relies on a general relationship between the two
congruences. Moreover, attempts to obtain a direct proof of the
completeness theorem for weak congruence which is simpler than the one
presented in \cite{AI95} have been to no avail.

\paragraph{Results}

The present paper extends the results from \cite{Fok94} and
\cite{AFGI95} from prefix iteration to {\em flat iteration}.  Flat
iteration was first mentioned in the technical report version of
\cite{BBP94}; it allows the first argument $P$ of $P^*Q$ to be a
(possibly empty) sum of actions.  For convenience, the CCS operator of
action-prefixing is also generalized to prefixing with sums of
actions.

My completeness proofs are considerably shorter than the
ones in \cite{AFGI95}. This is mostly a result of the presence of
expressions of the form $0^*P$ in the language, which allows a
collapse of several cases in the case distinction in \cite{AFGI95}.
In addition, the results for weak and delay congruence can be obtained
without using branching congruence as an intermediate step. Thanks to
the greater expressiveness of flat iteration, these results can be
reduced to the one for strong congruence, using the same proof
strategy as outlined above. However, the proposed reduction to strong
congruence does not work for $\eta$- and branching congruence.

In addition I derive the existing axiomatizations for prefix iteration
from the ones for flat iteration. In the case of weak congruence one
finds therefore that although a direct proof is cumbersome, there is a
choice between two attractive indirect proofs. One of them involves
first establishing the result for branching congruence; the other
involves first establishing the result for a richer language.

Finally, extending a result from \cite{BBP94}, I derive an expansion
theorem for the CCS parallel composition operator in the setting of
flat iteration. This is the key to extending the complete
axiomatizations of this paper to full CCS.  I show that such a theorem
does not exist in the setting of prefix iteration.

As in \cite{AFGI95}, my completeness proofs apply to open
terms directly, and thus yield\linebreak[2] the $\omega$-completeness of the
axiomatizations as well as their completeness for closed terms.
However, the generalization to full CCS applies to closed terms only.

\paragraph{Outline of the paper} Section~\ref{Sect:MPA-syntax}
introduces the language of basic CCS with flat iteration, $\LMPA$,
and its operational semantics. It also recalls the definitions of
strong, branching, $\eta$-, delay and weak congruence. The axiom
systems that will be shown to completely characterize the
aforementioned congruences over $\LMPA$ are presented in
Section~\ref{Sect:axiomatization}, and Section~\ref{Sect:completeness}
contains the proofs of their completeness.  In Section \ref{prefix}
the existing axiomatizations for prefix iteration are derived from the
ones for flat iteration.  Finally, Section \ref{parallelism} indicates
how the completeness results of this paper, unlike the ones for prefix
iteration, can, at least for closed terms, be extended to full CCS.

\section{Basic CCS with Flat Iteration}\label{Sect:MPA-syntax}

Assume a set $A$ of observable {\em actions}.  Let $\tau \not\in A$
denote a special {\em invisible action} and write $A_\tau := A \cup
\{\tau\}$. Also assume an infinite set $\Var$ of {\em variables},
disjoint with $A_\tau$.  Let $x,y,...$ range over $\Var$, $a,b,...$ over
$A$, $\alpha, \beta, \gamma, ...$ over $A_\tau$ and $\xi$ over $A_\tau
\cup \Var$.

The two-sorted language $\LMPA$ of basic CCS with flat iteration
is given by the BNF grammar:
\[
S :: = 0 \mid \alpha \mid S + S
\]
\[
P :: = x \mid 0 \mid S . P \mid P + P \mid S^*P
\]
Terms of sort $S$ are called {\em sumforms}, whereas terms of sort $P$
are called {\em process expressions}. The set of sumforms is denoted by
$\SF$ and the set of (open) process expressions by $\OMPA$. Let $s,t,u$
range over $\SF$ and $P,Q,R,S,T$ over $\OMPA$.  In writing terms over
the above syntax one may leave out redundant brackets, assuming that +
binds weaker than $.$ and $^*$.
For $I = \set{i_1 ,\ldots, i_n}$ a finite index set,
$\sum_{i\in I}P_i$ or $\sum\set{P_i \mid i\in I}$ denotes
$P_{i_1} + \cdots + P_{i_n}$.
By convention, $\sum_{i\in\emptyset}P_i$ stands for $0$.

The transition relations $\mv{\xi}$ are the least subsets of $(\OMPA
\times \OMPA) \cup \SF$ satisfying the rules in
Fig.~\ref{Fig:action-rules}. These determine the operational semantics
of $\LMPA$.  A transition $P \mv{\alpha} Q$ ($\alpha\in A_\tau$)
indicates that the system represented by the term $P$ can perform the
action $\alpha$, thereby evolving into $Q$, whereas $P \mv{x} P'$
means that the initial behaviour of $P$ may depend on the term that is
substituted for the process variable $x$.  It is not hard to see that
if $P\mv{x} P'$ then $P' = x$.  A transition $s \mv{\alpha}$ just says
that $\alpha$ is one of the actions in the sumform $s$.

\begin{figure}[htb]
\ThreeRules{}{\alpha \mv{\alpha}}
          {s \mv{\alpha}}{s + t \mv{\alpha}}
          {t \mv{\alpha}}{s + t \mv{\alpha}}
\FourRules{}{x \mv{x} x}
          {s \mv{\alpha}}{s . P \mv{\alpha} P}
          {P \mv{\xi} P'}{P + Q \mv{\xi} P'}
          {Q \mv{\xi} Q'}{P + Q \mv{\xi} Q'}
\TwoRules{s \mv{\alpha}}{s^*P \mv{\alpha} s^*P}
         {P \mv{\xi} P'}{s^*P \mv{\xi} P'}
\caption{\label{Fig:action-rules}Transition rules for $\LMPA$}
\end{figure}

\noindent
The set $\der{P}$ of {\em derivatives} of $P$ is the least set
containing $P$ that is closed under action-transitions.  Formally,
$\pder{P}$ is the least set satisfying:
\begin{center}
      if $Q \in \{P\}\cup\pder{P}$ and $Q \mv{\alpha} Q'$ for some
      $\alpha\in A_\tau$, then $Q'\in\pder{P}$,
\end{center}
and $\der{P} = \{P\} \cup \pder{P}$. Members of $\pder{P}$ are called
{\em proper} derivatives.

\begin{definition}\label{Def:bisimulation}\rm
Write $p \mvt{} q$ for $\exists n\!\geq\! 0\!: \exists p_0,...,p_n\!: p\!=\!p_0
\mv{\tau} p_1 \mv{\tau} ... \mv{\tau} p_n\! =\! q$, i.e.\ a
(possibly empty) path of $\tau$-steps from $p$ to $q$. Furthermore,
for $\xi \in A_\tau \cup \Var$, write ${p \mv{(\xi)} q}$ for $p
\mv{\xi} q \vee (\xi=\tau \wedge p=q)$. Thus ${\mv{(\xi)}}$
is the same as $\mv{\xi}$ for $\xi \in A\cup\Var$, and ${\mv{(\tau)}}$
denotes zero or one $\tau$-steps.

A {\em weak bisimulation} is a symmetric binary relation ${\cal R}$ on
$\OMPA$, such that 
\begin{equation}\label{bisimulation}
s{\cal R}t \wedge s\mv{\xi} s' \mbox{ implies } \exists
t_1,t_2,t': t\mvt{} t_1 \mv{(\xi)} t_2 \mvt{} t' \wedge s'{\cal R}t'.
\end{equation}
A weak bisimulation is a {\em delay bisimulation} if in the conclusion
of (\ref{bisimulation}) one has $t_2 = t'$. It is an
{\em $\eta$-bisimulation} if one has $s {\cal R} t_1$, and it is a
{\em branching bisimulation} if one has both $t_2 = t'$ and $s
{\cal R} t_1$. Finally, it is a {\em strong bisimulation} if one has
$$s{\cal R}t \wedge s\mv{\xi} s' \mbox{ implies } \exists
t': t \mv{\xi} t' \wedge s'{\cal R}t'.$$
Let $s,w,d,b$ be abbreviations for {\em strong}, {\em weak}, {\em
delay} and {\em branching}, and let $\wild$ range over
$\{s,w,d,\eta,b\}$. Then two processes $P,Q \in \OMPA$ are {\em
$\wild$(-bisimulation) equivalent}---notation $P \bis{\wild}{} Q$---if there
is a $\wild$-bisimulation $\cal R$ with $P {\cal R} Q$.
\end{definition}

\noindent
Following \cite{Mi89a,AFGI95}, the above definitions depart from the
standard approach followed in, {\eg}, \cite{Mi89} in that \pagebreak[3]
notions of bisimulation equivalence are defined that apply to open terms
directly. Usually, bisimulation equivalences like those presented in
Def.~\ref{Def:bisimulation} are defined explicitly for closed process
expressions only. Open process expressions are then regarded
equivalent iff they are equivalent under any closed substitution of
their (free) variables. In \cite{AFGI95} it has been shown, for the
language BCCS with prefix iteration, that both approaches yield the same
equivalence relation over open terms. The same proof applies to $\LMPA$.
For this result it is essential that the set $A$ of observable actions
is nonempty.

The following lemma will be of use in the completeness
proof for branching congruence (cf.~the proof of 
Propn.~\ref{Propn:prove-nf}). It is a standard
result for branching bisimulation equivalence.
\begin{lemma}[Stuttering Lemma \cite{GW96}]
\label{Lem.stuttering}
If $P_0\mv{\tau}\cdots\mv{\tau}P_n$ and $P_n\braeq P_0$, then
$P_i\braeq P_0$ for $i=1,...,n-1$.
\end{lemma}
The definition of $\bis{b}{}$ is equivalent to the one in \cite{GW96},
as follows immediately from the proof of the stuttering lemma in
\cite{GW96}. However, what is here introduced as a branching
bisimulation was there called a {\em semi branching bisimulation},
whereas ``branching bisimulation'' was the name of a slightly more
restrictive type of relation. The advantages of the current setup have
been pointed out in \cite{Bas95}.

\begin{proposition}\label{Propn:relations}
Each of the relations $\bis{\wild}{}$ {\em 
($\wild\in\set{s,b,\eta,d,w}$) }
is an equivalence relation and the largest $\wild$-bisimulation. 
Furthermore, for all $P,Q$, 
$$\begin{array}{ccccc}
P \bis{s}{} Q & \Rightarrow & P\braeq Q & \Rightarrow & P\deleq Q \\
&&\Downarrow && \Downarrow \\
&&  P\etaeq Q  & \Rightarrow &  P\obseq Q .
\end{array}$$
\end{proposition}
\begin{proof}
For $\wild\in\set{s,b,\eta,d,w}$, the identity relation, the converse of a 
$\wild$-bisimulation and the symmetric closure of the composition of two 
$\wild$-bisimulations are all $\wild$-bisimu\-la\-tions. Hence $\bis{\wild}{}$ 
is an equivalence relation. As pointed out in \cite{Bas95}, for
this argument to apply to branching bisimulations it is essential that
the definition of a branching bisimulation is relaxed to that of a
semi branching bisimulation.

That $\bis{\wild}{}$ is the largest $\wild$-bisimulation follows
immediately from the observation that the set of $\wild$-bisimulations
is closed under arbitrary unions. The implications hold by definition.
\end{proof}

For $s,t \in \SF$ write $s \leq t$ if $\forall \alpha (s \mv{\alpha}~
\Rightarrow t \mv{\alpha})$, and $s \bis{}{} t$ if $s \leq t$ and $t
\leq s$. It is easily checked that $\bis{}{}$ is a congruence on
sumforms in the sense that
$$\mbox{if } s\bis{}{}t \mbox{ then }
~s+u\bis{}{}t+u,~~u+s\bis{}{}u+t,~~s.P\bis{s}{}t.P,~ \mbox{
and }~s^* P\bis{s}{}t^* P.$$
Likewise, $\bis{s}{}$ turns out to be a
congruence on $\OMPA$ in the sense that
$$\mbox{if } P\bis{s}{}Q \mbox{ then }
~P+R\bis{s}{}Q+R,~~R+P\bis{s}{}R+Q,~~s.P\bis{s}{}s.Q~ \mbox{
and }~s^* P\bis{s}{}s^* Q.$$
However, for the standard reasons explained in, {\eg}, \cite{Mi89},
none of the equivalences $\bis{w}{}$, $\deleq$, $\etaeq$ and $\braeq$
is a congruence with respect to +. In fact, also none of these
equivalences is preserved by $^*$ \cite{AFGI95}. Following Milner
\cite{Mi89}, the solution to these congruence problems is by now
standard; it is sufficient to consider, for each equivalence
$\bis{\wild}{}$, the largest congruence over $\OMPA$ contained in
it. These largest congruences can be explicitly characterized as
follows.
\begin{definition}\label{Def:congruences}
\rm\mbox{ }
\begin{itemize}
\item $P$ and $Q$ are {\em branching congruent}, written $P \bracong Q$,
      iff for all $\xi\in A_\tau \cup \Vardef$,%
\vspace{-1ex}\begin{enumerate}
\item if $P \mv{\xi} P'$, then $Q \mv{\xi} Q'$ for some $Q'$ such that
      $P' \braeq Q'$;
\item if $Q \mv{\xi} Q'$, then $P \mv{\xi} P'$ for some $P'$ such that
      $P' \braeq Q'$.
\end{enumerate}
\item $P$ and $Q$ are {\em $\eta$-congruent}, written $P \etacong Q$,
      iff for all $\xi\in A_\tau \cup \Vardef$,
\vspace{-1ex}\begin{enumerate}
\item if $P \mv{\xi} P'$, then $Q \mv{\xi} Q_1 \mvt{} Q'$ for 
      some $Q_1, Q'$ such that $P' \etaeq Q'$;
\item if $Q \mv{\xi} Q'$, then $P \mv{\xi} P_1 \mvt{}P'$ for 
      some $P_1, P'$ such that $P' \etaeq Q'$.
\end{enumerate}
\item $P$ and $Q$ are {\em delay congruent}, written $P \delcong Q$,
      iff for all $\xi\in A_\tau \cup \Vardef$,
\vspace{-1ex}\begin{enumerate}
\item if $P \mv{\xi} P'$, then $Q \mvt{} Q_1 \mv{\xi}Q'$ 
      for some $Q_1, Q'$ such that $P' \deleq Q'$;
\item if $Q \mv{\xi} Q'$, then $P \mvt{} P_1 \mv{\xi} P'$ for 
      some $P_1, P'$ such that $P' \deleq Q'$.
\end{enumerate}
\item $P$ and $Q$ are {\em weakly congruent}, written $P \obscong Q$, 
      iff for all $\xi\in A_\tau \cup \Vardef$,
\vspace{-1ex}\begin{enumerate}
\item if $P \mv{\xi} P'$, then $Q \mvt{}\mv{\xi}\mvt{} Q'$ for some
      $Q'$ such that $P' \obseq Q'$;
\item if $Q \mv{\xi} Q'$, then $P \mvt{}\mv{\xi}\mvt{} P'$ for some
      $P'$ such that $P' \obseq Q'$.
\end{enumerate}
\item Finally, {\em strong congruence}, denoted $\bis{s}{c}$, is the same as
$\bis{s}{}$. 
\end{itemize}
\end{definition}
\begin{proposition}\label{Propn:largest}
For every $\wild\in\set{s,b,\eta,d,w}$, the relation $\bis{\wild}{c}$ is the 
largest congruence over $\OMPAdef$ contained in $\bis{\wild}{}$.
\end{proposition}
\begin{proof}
Exactly as in \cite{AFGI95}.
\end{proof}

\section{Axiom Systems}\label{Sect:axiomatization}

Table \ref{Tab:Fokkink} presents the axiom system ${\cal E}_s$, which
will be shown to completely characterize strong congruence over $\LMPA$.
The entries in this table are axiom schemes in the sense that
there is one axiom for every choice of the sumforms $s,t,u$.
For an axiom system ${\cal T}$, one
writes ${\cal T} \proves P = Q$ iff the equation $P = Q$ is provable
from the axiom system $\cal T$ using the rules of equational logic.
For a collection of equations $X$ over the signature of $\LMPAdef$,
$P \stackrel{\mbox{\tiny X}}{=} Q$ is used as a short-hand for 
$\mbox{\rm A1--A4,\it X} \proves P = Q$. 
The axioms A1--4 are known to completely characterize the operator + of CCS.
As this operator occurs both in sumforms and in process expressions,
these axioms appear for each of the two sorts. It is easily checked
that they are sound and complete for $\bis{}{}$ on sumforms:

\begin{table}[t]
\[
\begin{array}{|l rclcrcl |}
\hline
&&&&&&&\\
\mbox{\sc A1} & x + y & = & y + x & & s + t & = & t + s \\
\mbox{\sc A2} & (x + y) + z & = & x + (y + z)&& (s + t) + u & = & s + (t + u)\\
\mbox{\sc A3} & x + x & = & x && s + s & = & s \\
\mbox{\sc A4} & x + 0 & = & x && s + 0 & = & s \\
\mbox{\sc A5} &\multicolumn{3}{r}{(s+t).x} & =&\multicolumn{3}{l|}{s.x + t.x}\\
\mbox{\sc A6} &\multicolumn{3}{r}{0.x} & = &\multicolumn{3}{l|}{0}\\
\mbox{\sc FA1} &\multicolumn{3}{r}{0^*x} & = &\multicolumn{3}{l|}{x}\\
\mbox{\sc FA2} &\multicolumn{3}{r}{s^*(t.(s+t)^*x + x)} & = 
	      &\multicolumn{3}{l|}{(s+t)^*x} \\
&&&&&&&\\
\hline
\end{array}
\]
\caption{\label{Tab:Fokkink}The axiom system ${\cal E}_s$}
\vspace{-2ex}\end{table}

\begin{proposition}\label{Prop:sumforms}
$s \bis{}{} t \Leftrightarrow \mbox{\rm A1--4} \proves s=t$.
Moreover, $s \leq t \Leftrightarrow \mbox{\rm A1--4} \proves t=t+s$.
\end{proposition}
The axioms A5 and A6 are inspired by the ACP axioms for sequential
composition \cite{BK84}, and the axiom FA1 stems from \cite{Fok95},
where a form of iteration $P^*Q$ was used in which $P$ had to be
either an action, or a process (like 0) that cannot perform any actions.%
\pagebreak[3]
In \cite{BBP94} three axioms for general iteration in a process
algebra without 0 where proposed, called BKS1--3. These axioms where
shown to be complete in \cite{FZ94}. The axiom BKS2 deals with the
interaction between iteration and general sequential composition, and
therefore has no counterpart in $\LMPA$. My axiom FA2 is obtained from
BKS3 by requiring the first argument in an expression $P^*Q$ to be a
sumform. In the same spirit, the axiom BKS1 could be modified to
$t.(t^*x) + x = t^*x$. This law is derivable from ${\cal E}_s$ by setting
$s=0$ in FA2. The remaining axiom $a^*(a^*x) = a^*x$ of \cite{Fok94}
is derivable as well: take $s=t$ in FA2 and apply BKS1 to the
left-hand side.

\begin{table}[b]\vspace{-2ex}
\[
\begin{array}{|r|rlrcl@{~~~~}l|l|}
\hline
\multicolumn{1}{|l}{}&&&&&\multicolumn{3}{l|}{}\\
&\multicolumn{1}{|l|@{~}}{}
&\mbox{\sc FT1} & (s+\tau)^*x & = & \tau.(s^*x) + (s^*x) &
\multicolumn{2}{l|}{}\\
&\multicolumn{1}{|l|@{~}}{
   \raisebox{5pt}[0pt][0pt]{$~{\cal E}_b$}} &\mbox{\sc FT2} & \alpha.s^*(\tau.s^*(x+y) + x) & = & \alpha.s^*(x+y) &
\multicolumn{2}{l|}{}\\
~{\cal E}_\eta&&&&&\multicolumn{3}{l|}{}\\
&&\mbox{\sc T3} &\alpha.(x+\tau.y)&=&\alpha.(x+\tau.y)+\alpha.y &&\\
&&\mbox{\sc FT3} & s^* (x + \tau .y) & = & s^* (x + \tau .y + s.y) &&\\
\multicolumn{1}{|l}{}&&&&&&&{\cal E}_w~\\
\multicolumn{1}{|l}{} &&\mbox{\sc T1} & \alpha . \tau . x & = & \alpha
. x &\multicolumn{1}{|l|}{} &\\
\multicolumn{1}{|l}{}
&&\mbox{\sc FFIR} & (s+\tau)^*x & = & \tau.(s^*x) &\multicolumn{1}{|l|}{
   \raisebox{5pt}[0pt][0pt]{${\cal E}_d~$}} &\\
\multicolumn{1}{|l}{}&&&&&&\multicolumn{2}{l|}{}\\
\hline
\end{array}
\]
\caption{\label{Tab:equations-tau}Extra axioms for ${\cal E}_\eta$,
                 ${\cal E}_b$, ${\cal E}_d$ and ${\cal E}_w$}
\end{table}

In addition to the axioms in ${\cal E}_s$, the axiom systems ${\cal
E}_\wild$ ($\wild\in\set{b,\eta,d,w}$) include equations describing
the various ways in which the congruences $\bis{\wild}{c}$ abstract
away from internal actions $\tau$. These equations are presented
in Table~\ref{Tab:equations-tau}.  The axiom system
${\cal E}_b$ is obtained by adding the axioms FT1--2 to ${\cal E}_s$,
and ${\cal E}_\eta$ extends ${\cal E}_b$ with the equations T3 and
FT3. The set of axioms ${\cal E}_d$ consists of the axioms of ${\cal
E}_s$ together with T1 and FFIR.  Finally, ${\cal E}_w$ extends ${\cal
E}_d$ with T3 and FT3.

The equations T1 and T3 are standard laws for the silent action $\tau$
in weak congruence. Together with T2: $\tau.x = \tau.x + x$ and the
laws for strong congruence, they are known to completely characterize
weak congruence in the absence of iteration. Here T2 is derivable from
${\cal E}_s$ and FFIR (set $s=0$ in FFIR and apply BKS1 on
$\tau^*x$). Also the law $\alpha.(\tau.(x+y)+x) = \alpha.(x+y)$, which
together with the laws for strong congruence characterizes branching
bisimulation for BCCS without iteration, is derivable: just take
$s=0$ in FT2.

The four remaining axioms, which describe the interplay between $\tau$
and prefix iteration, are new here.  The law FFIR is a generalization
of the {\em Fair Iteration Rule} $\tau^*x = \tau.x$ (FIR$_1$) of
\cite{BBP94}, which is an equational formulation of {\sl Koomen's Fair
Abstraction Rule} \cite{BBK87a}. Like FIR, FFIR expresses that modulo
weak (or delay) congruence a process remains the same if $\tau$-loops
are added (or deleted) in (or from) its proper derivatives. The law
FT1 has the same function in branching (or $\eta$-)bisimulation
semantics, but has to be formulated more carefully because T2 is not
valid there. Note that FT1 can be reformulated as $\alpha.(s+\tau)^*x =
\alpha.s^*x$.  The laws FT2 and FT3 are straightforward generalizations of the
laws PB2 and PT3 of \cite{AFGI95}.
The remaining law PT2 of \cite{AFGI95} is (by the forthcoming
completeness theorem for $\delcong$) derivable from the ones given here.

Note that even over a finite alphabet $A$ there exist infinitely many
sumforms. Hence the axiomatizations as given here are infinite.
However, for each axiom scheme only the instantiations are needed in
which the sumforms have the form $\sum_{i=1}^n \alpha_i$ in which all
the $\alpha_i$'s are different. With this modification each of the
axiom systems ${\cal E}_\wild$ ($\wild\in\set{s,b,\eta,d,w}$) is
finite if so is the set of actions $A$.  If $A$ is not finite, the
axiomatizations can still be interpreted as finite ones, namely by
replacing the actions $\alpha$ in FT2 and T2,3 by sumforms $t$,
introducing variables that range over sumforms, and interpreting each
entry in the resulting Tables~1--4 as a single axiom in which $s$, $t$
and $u$ are such variables.

The following states the soundness of the axiom systems.
\begin{proposition}\label{Propn:soundness}
Let $\wild\in\set{s,b,\eta,d,w}$.
If ${\cal E}_\wild \proves P = Q $, then $P \bis{\wild}{c} Q$.
\end{proposition}
\begin{proof}
  As $\bis{\wild}{c}$ is a congruence, it
  is sufficient to show that each equation in ${\cal E}_\wild$ is
  sound with respect to it. This is rather straightforward and left to
  the reader.
\end{proof}

As in \cite{AFGI95}, it can be shown that
${\cal E}_w \proves {\cal E}_d \proves {\cal E}_b \proves {\cal E}_s$ and
${\cal E}_w \proves {\cal E}_\eta \proves {\cal E}_b$, where
${\cal T} \proves {\cal T}'$ denotes that ${\cal T} \proves P = Q$ for
every equation $(P=Q)\in{\cal T}'$.

\section{Completeness}\label{Sect:completeness}

This section is entirely devoted to detailed proofs of the
completeness of the axiom systems ${\cal E}_\wild$
($\wild\in\set{s,b,\eta,d,w}$) with respect to $\bis{\wild}{c}$ over the
language of open terms $\OMPA$. The first subsection contains the
completeness proof for branching congruence. Its contents also apply
to strong congruence if you read ${\cal E}_s$ for ${\cal E}_b$,
$\bis{s}{}$ for $\braeq$, $\alpha$ for $a$, and $\alpha$ for
$(\alpha)$ and skip the underlined  and sidelined parts.

\subsection{Completeness for strong \bran{and branching} congruence}

First I identify a subset of process expressions of a special
form, which will be convenient in the proof of the completeness
result. Following a long-established tradition in the literature on
process theory, these terms are referred to as {\sl normal
forms}. The set of normal forms is the smallest set of
process expressions of the form
\[
s^*(\sum_{i\in I}\alpha_i.P_i+\sum_{j\in J} x_j),
\]
where \bran{$s \nv{\tau}$,} the terms $P_i$ are themselves normal forms, and
$I,J$ are finite index sets. (Recall that the empty sum represents 0.)
\begin{lemma}
\label{Lem:normalforms}
Each term in $\OMPAdef$ can be proven equal to a normal form using equations 
{\em A1--6}, {\em FA1,2} \bran{and {\rm FT1}}.
\end{lemma}
\begin{proof}
A straightforward induction on the structure of process expressions.
The expressions $x$ and $0$ can be brought in the required form by a
single application of FA1. Now suppose $P$ and $Q$ have the required
form. Then $s.P$ can be brought in normal form using A5 or A6
(possibly after applying A4 on $s$), followed by FA1. $P+Q$ can be
brought in normal form by first applying the derivable law $t^*x =
t.(t^*x) + x$ (BKS1) on each of $P$ and $Q$, then A4--6 to rewrite the
subterms $t.(t^*x)$, and concluding with FA1. Finally $s^*P$ is dealt
with by applying BKS1 on $P$, again followed by A4--6. \bran{In case $s
\mv{\tau}$, apply FT1, followed by another round of BKS1, A4--6 and FA1.}
\end{proof}

\noindent
Note that this is the only place in the completeness proof where the
axioms FA1 \bran{and FT1} are used.
The following result is \bran{the key to} the completeness theorem.
\begin{proposition}\label{Propn:prove-nf}
For all $P,Q\in\OMPAdef$, if $P \braeq Q$, then, $\bran{\forall
\gamma\in A_\tau:}
{\cal E}_b \proves \bran{\gamma.}P = \bran{\gamma.}Q$.
\end{proposition}
\begin{proof}
First of all, note that, as the equations in ${\cal E}_b$ are sound with 
respect to $\bracong$, and, {\sl a fortiori}, with respect to $\braeq$, by 
Lem.~\ref{Lem:normalforms} it is sufficient to prove that the statement of
the proposition holds for \bran{branching} equivalent normal forms $P$ and $Q$.
I do so by complete induction on the sum of the sizes of $P$ and $Q$.

Let $P\ACeq s^*(\sum_i\alpha_i.P_i+\sum_k x_k)$ and
$Q\ACeq t^*(\sum_j\beta_j.Q_j+\sum_l y_l)$.
Write $P'$ for $\sum_i\alpha_i.P_i+\sum_k x_k$ and
$Q'$ for $\sum_j\beta_j.Q_j+\sum_l y_l$.
Consider the following two conditions:
\begin{itemize}
\item[A.]
$P_i\braeq Q$ for some $i$;
\item[B.]
$Q_j\braeq P$ for some $j$.
\end{itemize}
I distinguish two cases in the proof, depending on which of these
conditions hold.
\begin{itemize}
\item[I]
Suppose that both of A and B hold. In this case, there exist $i$ and $j$
such that $P_i \braeq Q \braeq P \braeq Q_j$.  Applying the inductive
hypothesis to the equivalences $P \braeq Q_j$, $Q_j \braeq P_i$ and
$P_i \braeq Q$, one infers that\bran{, for all $\gamma\in A_\tau$,}
\[
{\cal E}_b \proves \bran{\gamma.}P = \bran{\gamma.} Q_j = \bran{\gamma.} P_i = \bran{\gamma.} Q
\]
\item[II]
Suppose that at most one of A and B holds. 
Assume, without loss of generality, that B does not hold.

Suppose $s \mv{a}$. As $P \braeq Q$, the transition 
$P\mv{a}P$ must be matched by a sequence of transitions
$Q\bran{=Q_0 \mv{\tau} Q_1 \mv{\tau} \cdots \mv{\tau} Q_n} \mv{a}Q''$
with \bran{$P \braeq Q_n$ and} $P\braeq Q''$. As condition B does
{\sl not} hold, \bran{using Lem.~\ref{Lem.stuttering}} it follows that
\bran{$n=0$,} $Q'' = Q$ and $t \mv{a}$. Hence ${\cal E}_b \proves t=t+s$ by
Prop.~\ref{Prop:sumforms}.

Let $u \ACeq \sum\set{\alpha_i \mid  P_i \bis{b}{} Q \wedge ( t
\mv{\alpha_i} \bran{\vee \alpha_i = \tau})}$ and $v \ACeq \sum\set{\alpha_i \mid  P_i
\bis{b}{} Q \wedge t \mv{\alpha_i}}$. Then ${\cal E}_b \proves t=t+v=t+s+v$.

For every summand $\alpha_i.P_i$ of $P'$ with $P_i\braeq Q$,
induction yields ${\cal E}_b\proves\alpha_i.P_i=\alpha_i.Q$.
Hence, using axiom A5 to assemble all such summands with $u
\mv{\alpha_i}$, and possibly using A4 and/or A6 if there are no or
only such summands, one infers that
\[
{\cal E}_b \proves P=s^*(u.Q+S) 
\]
where $S\ACeq \sum\set{\alpha_i . P_i \mid  P_i \nobis{b}{} Q \vee (
t \nv{\alpha_i} \bran{\wedge \alpha_i \neq \tau} )} + \sum_k x_k$.

Consider now a summand $\alpha_i.P_i$ of $S$. 
As $P \braeq Q$, the transition 
$P\mv{\alpha_i}P_i$ must be matched by a sequence
$Q\bran{=Q_0 \mv{\tau} Q_1 \mv{\tau} \cdots \mv{\tau} Q_n} \mv{(\alpha_i)}Q''$
with \bran{$P \braeq Q_n$ and} $P_i\braeq Q''$. \bran{As condition B does
not hold, using Lem.~\ref{Lem.stuttering} it follows that $n=0$.}
Furthermore, the possibility $Q \mv{(\alpha_i)} Q \braeq P_i$ is
ruled out by the construction of $S$.
Hence, each summand $\alpha_i.P_i$ of $S$ 
matches with a summand $\beta_j.Q_j$ of $Q'$, in the sense that
$\alpha_i=\beta_j$ and $P_i\braeq Q_j$. For each such pair of related
summands, induction yields 
\[
{\cal E}_b \proves \alpha_i.P_i=\alpha_i.Q_j=\beta_j.Q_j \enspace .
\]
Moreover, each summand $x_k$ of $S$ must be a summand of $Q'$. Hence, possibly
using axiom A3, it follows that ${\cal E}_b \proves Q' = Q'\! + S$.
\bran{Now I distinguish two sub-cases.}
\item[IIa]
\bran{Suppose that A does not hold for an index $i$ with 
$\alpha_{i} \hspace{-.15pt} = \hspace{-.15pt} \tau$.}
Again \bran{using Lem.~\ref{Lem.stuttering},} it follows that every summand
$\beta_j.Q_j$ of $Q'$ matches with a summand $\alpha_i.P_i$ of $S$
(since also B does not hold, the cases $Q_j \braeq P$ and $Q_j \braeq
Q \braeq P$ do not apply), and every $y_l$ is equal to an
$x_k$. Possibly using axiom A3, it follows that ${\cal E}_b \proves S=
Q' + S = Q'$. Moreover, whenever $t \mv{a}$ then $Q \mv{a} Q$, so $P
\mv{a} P'' \braeq Q$ and hence either $s \mv{a}$ or $v \mv{a}$.
It follows that ${\cal E}_b \proves t=s+v$. Finally $u=v$, so
\[
\bran{\gamma.}P=\bran{\gamma.}s^*(v.Q+S)=\bran{\gamma.}s^*(v.(s+v)^*S+S)
\stackrel{\mbox{\tiny \rm FA2}}=\bran{\gamma.}(s+v)^*S=\bran{\gamma.}Q.
\]
\item[IIb]
Suppose that A holds for an index $i$ with $\alpha_{i}=\tau$.
Then ${\cal E}_b \proves u=\tau +v$, so
$$
\begin{array}{@{}r@{~}c@{~}l@{}}
\gamma.P&\stackrel{\mbox{\tiny \rm A5}}=&\gamma.s^*(\tau.Q+v.Q+S)\\
&=&\gamma.s^*\Parens{\tau.t^*\Parens{Q'+S}+v.t^*\Parens{Q'+S}+S}\\
&\stackrel{\mbox{\tiny \rm FA2}}=&\gamma.s^*\Parens{\tau.s^*\Parens{t.(s+t)^*\Parens{Q'+S}+Q'+S}+v.t^*\Parens{Q'+S}+S}\\
&\stackrel{\mbox{\tiny \rm ~}}=&\gamma.s^*\Parens{\tau.s^*\Parens{Q'+(t+v).t^*\Parens{Q'+S}+S}+v.t^*\Parens{Q'+S}+S}\\
&\stackrel{\mbox{\tiny\rm FT2, A5}}=&\gamma.s^*\Parens{Q'+(t+v).t^*\Parens{Q'+S}+S}\\
&\stackrel{\mbox{\tiny \rm FA2}}=&\gamma.t^*(Q'+S)~=~\gamma.Q.
\makebox[0pt][l]{\hspace{2.5in}\raisebox{0pt}[0pt][0pt]{\rule{.3pt}{1.7in}}}
\end{array}
$$
\end{itemize}
The proof of the inductive step is now complete.
\end{proof}

\begin{theorem}
\label{Thm:branching-completeness}
Let $P,Q\in\OMPAdef$. 
If $P \bracong Q$, then ${\cal E}_b \proves P = Q$.
\end{theorem}
\begin{proof}
Consider two process expressions $P$ and $Q$ that are branching
congruent. Using the same technique as in the proof of
Lem.~\ref{Lem:normalforms}, one may derive that
\begin{center}$\begin{array}{l}
{\cal E}_b \proves P = \sum\set{ \alpha_i. P_i \mid i \in I} + 
                       \sum\set{ x_j \mid j\in J} 
                       ~~~\mbox{and} \\
{\cal E}_b \proves Q = \sum\set{ \beta_k. Q_k \mid k \in K} + 
                       \sum\set{ y_l \mid l\in L} 
\makebox[0pt][l]{\hspace{1.3in}\raisebox{0pt}[0pt][0pt]{\rule{.3pt}{.9in}}}
\end{array}$\vspace{-2ex}\end{center}
\pagebreak[3]
for some finite index sets $I, J, K, L$. 
As $P\bracong Q$, it follows that
\begin{enumerate}
\item for every $i \in I$ there exists an index $k_i \in K$ such that
  $\alpha_{i} = \beta_{k_i}$ and $P_{i} \braeq Q_{k_i}$,
\item and for every $j \in J$ there exists an index $l_j \in L$ such
  that $x_{j} = y_{l_j}$.
\end{enumerate}
By Propn.~\ref{Propn:prove-nf}, for every $i \in I$ one may infer that
\[
{\cal E}_b \proves \alpha_{i}.P_{i} = \alpha_{i}.Q_{k_i} = \beta_{k_i}.Q_{k_i} 
\enspace .
\] 
Using A3 it follows immediately that ${\cal E}_b \proves Q = P+Q$.
By symmetry one obtains ${\cal E}_b \proves P = P+Q = Q$.
\marginpar[t]{\hspace{-3mm}\rule{.3pt}{1.43in}}
\end{proof}

\subsection{Completeness for $\eta$-, delay, and weak congruence}

I now proceed to derive completeness results for $\eta$-, delay, and 
weak congruence from the ones for strong and branching congruence. The key
to this derivation is the observation that, for certain classes of process
expressions, these congruence relations coincide with $\bis{s}{}$ or
$\bis{b}{c}$. These classes of process expressions are defined below.
\begin{definition}\label{Def:saturated}\rm
A term $P$ is:
\begin{itemize}
\item {\em $\eta$-saturated} iff for each of its derivatives $Q,~R$ and $S$ 
      and $\xi \in A_\tau \cup \Vardef$ one has that:
      \[
      Q \mv{\xi} R \mv{\tau} S \mbox{ implies }  Q \mv{\xi} S.
      \]
\item {\em $d$-saturated} iff  for each of its derivatives $Q,~R$ and $S$ 
      and $\xi \in A_\tau \cup \Vardef$ one has that:
      \[
      Q \mv{\tau} R \mv{\xi} S \mbox{ implies }  Q \mv{\xi} S.
      \]
\item {\em $w$-saturated} iff it is both $\eta$- and $d$-saturated.
\item {\em strongly $\wild$-saturated} (for $\wild \in \{\eta,d,w\}$)
      if it is $\wild$-saturated and for each of its proper
      derivatives $Q \in \pder{P}$ there is a {\em $\tau$-loop} $Q\mv{\tau} Q$.
\end{itemize}
\end{definition}
The following was first observed in \cite{GW96} for process graphs. 

\begin{theorem}\label{Thm:saturated}\mbox{ }
\begin{list}{$\bullet$}{\leftmargin 8pt
\topsep 4pt \itemsep 2pt \parsep 2pt}
\item [1.]
      If $P$ and $Q$ are $\wild$-saturated, $\wild\in\set{\eta,d,w}$,
      and $P \bis{\wild}{c} Q$, then $P \bracong Q$. 
\item [2.]
      \mbox{If $P$ and $Q$ are strongly $\wild$-saturated,
      $\wild\!\in\!\set{d,w}$, and $P \bis{\wild}{c} Q$, then $P \bis{s}{} Q$.}
\end{list}
\end{theorem}
\begin{proof}
In case 1, the relation
\begin{eqnarray*}
\Rel{B} & \deq & \set{ (S,T) \mid S \bis{\wild}{} T,~~\mbox{$S,T$ $\wild$-saturated} }
\end{eqnarray*} 
is a branching bisimulation. From this it follows easily (as shown in
\cite{AFGI95}) that $P \bis{\wild}{c} Q$ implies $P \bracong Q$. 
In case 2, $\Rel{B}$ is a strong bisimulation.
\end{proof}

\noindent
Note that the second statement does not apply to $\etacong$. A
counterexample concerns the terms $P=a.\tau^*\tau.\tau^*b.\tau^*0 +
a.\tau^*b.\tau^*0$ and $Q=a.\tau^*b.\tau^*0$. These terms are strongly
$\eta$-saturated and $P \etacong Q$, but $P \nobis{s}{} Q$.

\begin{theorem}\label{saturated}
Let $\wild\in\set{\eta,d,w}$. 
\begin{list}{$\bullet$}{\leftmargin 8pt
\topsep 4pt \itemsep 2pt \parsep 2pt}
\item [1.]
For each term $P$,
${\cal E}_\wild \proves P = P'$ for some $\wild$-saturated
term $P'$.
\item [2.]
For each term $P$,
${\cal E}_\wild \proves P = P''$ for some strongly $\wild$-saturated
term $P''$.
\end{list}
\end{theorem}
\begin{proof}
The first statement has been shown in \cite{AFGI95} for the language
BCCS$^{p*}$. The resulting term $P'$ has the form 
$P' = \sum_{i\in I} \alpha_i . P_i + \sum_{j\in J} x_j $.
The same proof applies here.

For the second result, first prove $P$ equal to a term $P'$ as above,
and bring the subterms $P_i$ for $i \in I$ in normal form, using
Lem.~\ref{Lem:normalforms}. Now each proper derivative of the
resulting term has the form $s^*Q$, and appears in a subterm of the form
$\alpha.s^*Q$. In combination with T1, the axiom FFIR derives
$\alpha.s^*x = \alpha.(s+\tau)^*x$. As mentioned before, this law is
also derivable from ${\cal E}_b$. Applying this law to all subterms of
the form $\alpha.s^*Q$ results in a term $P''$ that is still
$\wild$-saturated, and for which each proper derivative $Q$ has a
$\tau$-loop $Q \mv{\tau} Q$.
\end{proof}

\noindent 
The results in 
Thms.~\ref{Thm:saturated}.1 and~\ref{saturated}.1 effectively 
reduce the completeness problem for $\eta$-, delay, and weak
congruence over $\OMPA$ to that for branching congruence.

\begin{corollary}\label{Cor:other-completeness} 
Let $\wild\in\set{\eta,d,w}$. If
$P\bis{\wild}{c}Q$, then ${\cal E}_\wild \proves P = Q$.
\end{corollary}
{\bf Proof (for the case $\wild=\eta$):} Suppose that $P \bis{\wild}{c} Q$. 
Prove $P$ and $Q$ equal to $\wild$-saturated processes $P'$ and $Q'$, 
respectively (Thm.~\ref{saturated}.1). By the soundness of the 
axiom system ${\cal E}_\wild$ (Propn.~\ref{Propn:soundness}), $P'$ and $Q'$ 
are $\wild$-congruent. It follows that $P'$ and $Q'$ are branching
congruent (Thm.~\ref{Thm:saturated}.1). 
Hence, by Thm.~\ref{Thm:branching-completeness}, 
${\cal E}_b\proves P' = Q'$. The claim now follows because
${\cal E}_b \subset {\cal E}_\eta$.\hfill $\Box$\vspace{2ex}

\noindent
The cases $\wild = d$ and $\wild = w$ can be proved in the same way,
using in the last step that ${\cal E}_w \proves {\cal E}_d \proves
{\cal E}_b \proves P=Q$ (cf.~the last sentence of
Section~\ref{Sect:axiomatization}). However, 
Thms.~\ref{Thm:saturated}.2 and~\ref{saturated}.2 allow a
simpler proof that doesn't need the completeness result for branching
bisimulation as an intermediate step, but instead
reduces the problem to the completeness for strong congruence.
\vspace{2ex}

\noindent
{\bf Proof of Corollary~\ref{Cor:other-completeness} (for the cases
$\wild \in \{d,w\}$):} Suppose that $P \bis{\wild}{c} Q$. 
Prove $P$ and $Q$ equal to strongly $\wild$-saturated processes $P'$ and $Q'$, 
respectively (Thm.~\ref{saturated}.2). By the soundness of the 
axiom system ${\cal E}_\wild$ (Propn.~\ref{Propn:soundness}), $P'$ and $Q'$ 
are $\wild$-congruent. It follows that $P'$ and $Q'$ are strong
congruent (Thm.~\ref{Thm:saturated}.2). 
Hence, by Prop.~\ref{Propn:prove-nf} for strong congruence, 
${\cal E}_s\proves P' = Q'$. The claim now follows because
${\cal E}_s \subset {\cal E}_\wild$.\hfill $\Box$

\section{Prefix Iteration}\label{prefix}

In this section I derive complete axiomatizations for prefix iteration
from the ones for flat iteration. A $\LMPA$ process expression is
a BCCS$^{p*}$ expression iff in each subexpression $s.P$ or $s^*P$,
the sumform $s$ consists of a single action $\alpha \in A_\tau$.
The following result about the expressiveness of
BCCS$^{p*}$ stems from \cite{AI95}.

\begin{lemma}
\label{lem.Luca}
If $P_0$ is a {\rm BCCS}$^{p*}$ expression and
$P_n\mvt{}\mv{a_n}P_{n+1}$ for $n=0,1,2,...$, then there
is an $N$ such that $a_n=a_N$ for $n>N$.
\end{lemma}

\begin{definition}\label{Def:potential}\rm
A $\LMPA$ expression $P_0$ is a {\em potential {\rm BCCS}$^{p*}$ expression}
if every sequence $P_n\mvt{}\mv{a_n}P_{n+1}$ ($n=0,1,2,...$) has the
property of Lem.~\ref{lem.Luca}.
\end{definition}
It is easy to see that a potential BCCS$^{p*}$ expression can not be
weakly equivalent to an expression that is not so. Hence, using
Propn.~\ref{Propn:soundness} (soundness):

\begin{lemma}\label{Lem:both}
Let $\wild \in \{s,b,\eta,d,w\}$. If ${\cal E}_\wild \proves P=Q$ then
either both $P$ and $Q$ are potential {\rm BCCS}$^{p*}$ expressions,
or neither of them is.
\end{lemma}
Using structural induction, the following Lemma is
straightforward:

\begin{lemma}\label{Lem:potential}
If $s^*P$ is a subterm of a potential {\rm BCCS}$^{p*}$ expression, then
either $\mbox{\rm A1--4} \proves s=0$ or $\mbox{\rm A1--4} \proves
s=\alpha \in A_\tau$ or $\mbox{\rm A1--4} \proves s=a+\tau$ with $a
\in A$. Moreover, these alternatives are mutually exclusive.
\end{lemma}
Let $R$ be the rewrite system consisting of the axioms A5, A6, FA1
and FT1, read from left to right. As these rewrite rules have no
overlapping redexes, $R$ is confluent, and it is equally
straightforward to see that it is terminating. Now let $\phi$ be the
operator on potential BCCS$^{p*}$ expressions $P$ that first converts any
sumform $s$ in a subterm $s^*Q$ of $P$ into one of the forms $0$,
$\alpha$ or $a+\tau$ (using A1--4 and Lem.~\ref{Lem:potential}), and
subsequently brings the resulting term in normal form w.r.t.\ $R$.
Note that the resulting term $\phi(P)$ is a BCCS$^{p*}$ expression.

\begin{theorem}
Let $\wild \in \{b,\eta,d,w\}$. The theory $$\phi({\cal E}_\wild) =
\{ \phi(P) = \phi(Q) \mid (P=Q) \in {\cal E}_\wild \}$$ is a complete
axiomatization of $\bis{\wild}{c}$ over the language {\rm BCCS}$^{p*}$.
\end{theorem}
\begin{proof}
An equation $P=Q$ is provable in equational logic iff
there exists a sequence $T_0, ... ,T_n$ with $P=T_0$, $Q=T_n$, and
the equation $T_{i-1} = T_i$ is obtained from one axiom by means of
substitution, placement in a context and (possibly) symmetry
($i=1,...,n$). 
Suppose that $P \bis{\wild}{c} Q$ for certain
BCCS$^{p*}$ expressions $P$ and $Q$. As $P$ and $Q$ are also
BCCS$^{f*}$ expressions, this implies ${\cal E}_\wild \proves P=Q$.
Thus, by Lem.~\ref{Lem:both}, there exists a proof-sequence as
mentioned above in which all the $T_i$ are potential BCCS$^{p*}$
expressions. Now, for $i=1,...,n$, the equation $\phi(T_{i-1}) =
\phi(T_i)$ can be obtained from an axiom in $\phi({\cal E}_\wild)$ by
means of substitution, placement in a context and symmetry.  This
yields a proof-sequence for the equation $\phi(P)=\phi(Q)$.  However,
since $P$ and $Q$ are BCCS$^{p*}$ expressions, $\phi(P)=P$ and
$\phi(Q)=Q$. Hence $\phi({\cal E}_\wild) \proves P=Q$.
\end{proof}

\noindent
In the axiom systems $\phi({\cal E}_\wild)$, the axioms A5, A6 and FA1
evaluate to identities, whereas the axioms A1--4, T1 and T3 remain
unchanged.  Furthermore, there are three axioms corresponding to each
of FT1--3 and FFIR, depending on whether $s$ evaluates to $0$,
$\alpha$, or $a+\tau$, and nine corresponding to FA2, depending on how
$s$ and $t$ evaluate. All resulting axiomatizations turn out to be
derivable from the corresponding axiomatizations in \cite{AFGI95} and
vice versa. Hence the above constitutes an alternative proof of the
completeness results in \cite{AFGI95}.

A similar result can be obtained for $\wild = s$, but in that case
$\tau$ should be treated as a normal action, and FT1 should be omitted
from the rewrite system.

\section{Parallelism}\label{parallelism}

Complete axiomatizations of strong and weak bisimulation congruence
over full CCS without recursion or iteration were given in
\cite{HM85}. The strategy, in both cases, was to prove every such CCS
expression strongly equivalent to a BCCS expression, using the well
known {\em expansion theorem}, and then apply the relevant
completeness theorem for BCCS expressions. This method does not work
in the setting of prefix iteration, as the parallel composition of two
BCCS$^{p*}$ expressions need not be (weakly) equivalent to a BCCS$^{p*}$
expression. A simple counterexample concerns the expression $a^*0 \mid
b^*0$, which is not a potential BCCS$^{p*}$ expression in the sense of
Def.~\ref{Def:potential}. However, an expansion theorem for CCS$^{f*}$
poses no problem: 
let $P= s^* \sum_{i\in I} \alpha_i. P_i$ and $Q = t^* \sum_{j\in J}
\beta_j. Q_j$, then $$P \mid Q \bis{s}{} (s+t+\gamma)^* 
\Parens{\displaystyle\sum_{i\in I}
\alpha_i. (P_i \mid Q) + \sum_{j\in J} \beta_j. (P \mid Q_j) + C}\vspace{-2ex}$$
with $$C = \sum_{\alpha_i = \overline{\beta_j}} \tau. (P_i \mid Q_j)
+ \sum_{i \in I,~ t\mv{\overline{\alpha_i}}} \tau.(P_i \mid Q)
+ \sum_{j \in J,~ s\mv{\overline{\beta_j}}} \tau.(P \mid Q_j)$$
and $\gamma = \left\{\begin{array}{ll}
   \tau & \mbox{if there is an $a \in A$ with $s \mv{a}$ and $t \mv{a}$} \\
   0    & \mbox{otherwise.}\end{array}\right.$
\\[2ex]
For a parallel composition without communication just leave out
$\gamma$ and $C$; in this shape the theorem was first found in \cite{BBP94}.

In the presence of a CSP-style parallel composition in which
processes are forced to synchronize over a shared alphabet
\cite{OH86}, closed expressions with flat iteration can be expressed
in terms of prefix iteration. An expression $$(a+b)^*(c.P+d.Q)$$ for
instance, in which $c$ and $d$ do not occur in $P$ and $Q$, is
strongly equivalent to $$a^*(c.0+d.0) \|_{\{c,d\}} b^*(c.P+d.Q)$$ where
synchronization over $c$ and $d$ is enforced. In the general case
renaming operators are needed as well.

\paragraph{Acknowledgment}
The inspiration to write this paper originated from the fruitful
collaboration with Luca Aceto, Wan Fokkink and Anna
Ing\'{o}lfsd\'{o}ttir in \cite{AFGI95}. The referees are thanked for
careful proofreading, and correcting the expansion theorem above.


\end{document}